\documentclass[english,aps,pre,twocolumn,showpacs,amsmath,amssymb]{revtex4-1}

\usepackage{color}
\usepackage{graphicx}
\usepackage{amsmath}
\usepackage{placeins}

\begin{document}

\title{Confined Sandpile in two Dimensions: Percolation and Singular Diffusion}
\author{R.~S.~Pires}
\author{A.~A.~Moreira}
\author{H.~A.~Carmona}
\author{J.~S.~Andrade~Jr.}

\affiliation{Departamento de F\'{\i}sica, Universidade Federal do
 Cear\'a, 60451-970 Fortaleza, Cear\'a, Brazil}

\date{\today}

\begin{abstract}
We investigate the properties of a two-state sandpile model subjected
to a confining potential in two dimensions. From the microdynamical
description, we derive a diffusion equation, and find a stationary
solution for the case of a parabolic confining potential. By studying the
systems at different confining conditions, we observe two scale-invariant
regimes. At a given confining potential strength, the cluster size
distribution takes the form of a power law. This regime
corresponds to the situation in which the density at the center of the system approaches the
critical percolation threshold. The analysis of the fractal dimension of
the largest cluster frontier provides evidence that this regime is
reminiscent of gradient percolation. By increasing further the confining
potential, most of the particles coalesce in a giant cluster, and we
observe a regime where the jump size distribution takes the form of a
power law. The onset of this second regime is signaled by a maximum
in the fluctuation of energy.
\end{abstract}
\maketitle

\section{Introduction}

Anomalous diffusion is
observed in many physical scenarios such as fluid transport in porous
media~\cite{Kuntz2001,Lukyanov2012}, diffusion in crowded fluids~\cite{Szymanski2009},
diffusion in fractal-like
substrates~\cite{Stephenson1995,Andrade1997,Buldyrev2001,Costa2003,Havlin2002},
turbulent diffusion in the atmosphere~\cite{Richardson1926,Hentschel1984},
diffusion of proteins due to molecular crowding~\cite{Banks2005}, systems
including ultra-cold atoms~\cite{Sagi2012}, analysis of heartbeat
histograms~\cite{Grigolini2001}, diffusion in
``living polymers''~\cite{Ott1990} and study of financial
transactions~\cite{Plerou2000}. Anomalous diffusion can also
manifest its non-Gaussian behavior in terms of nonlinear Fokker-Plank
equations~\cite{Lenzi2001,Malacarne2001,Malacarne2002,DaSilva2004,Lenzi2005},
which is the case, for example, of the dynamics of interacting vortices in
disordered superconductors~\cite{Zapperi2001,Moreira2002,Miguel2003,Andrade2010},
diffusion in dusty plasma~\cite{Liu2008,Barrozo2009}, and pedestrian
motion~\cite{Barrozo2009}. A very interesting case of anomalous diffusion
is surely the singular diffusion which is identified as having a divergent
diffusion coefficient~\cite{Carlson1990,Carlson1993,Carlson1995,Kadanoff1992,Barbu2010}. This kind of behavior happens in
nature in some physical situations, for instance when adsorbates diffuse on a
adsorbent surface, its diffusion can be very nonlinear with a diffusion
coefficient which depends on the local coverage $\theta$ as,
$D\propto{\left|\theta-\theta_{c}\right|^{-\alpha}}$~\cite{Myshlyavtsev1995}.
Therefore, the study of the basic mechanisms behind surface diffusion is of large importance for
understanding technologically important processes like physical
adsorption~\cite{Vidali1991} and catalytic surface
reactions~\cite{Manandhar2003,Hofmann2005,Schmidtbauer2012}.

A special class of singular diffusion models were intensively studied by
Carlson~\textit{et~al.}~\cite{Carlson1990,Carlson1993} in two state 1D sandpile
models for which they derive diffusion equations with singularities in the
diffusion coefficient of the form $D\propto(1+\rho)/(1-\rho)^3$, where $\rho$ is
the local density. They suggest that some open driven systems present
self-organized criticality~\cite{Bak1987} because in their continuum limit singularities appear
in the diffusion constant at a critical point~\cite{Carlson1990}. Some
characteristics of this model change drastically when a confining potential is
applied~\cite{Pires2015}. The jump-size distribution, for instance, starts to
exhibit a power-law behavior which suggests a scale-invariant behavior of the
system~\cite{Pires2015}. Scale-invariant behavior in diffusive systems were also observed in gradient
percolation diffusion fronts in 2D~\cite{Sapoval1985}, that have been shown to
display fractal diffusion fronts with characteristic dimension similar to the
boundary of critical percolation clusters~\cite{Sapoval1985}.

In this paper we investigate a 2D confined sandpile model. Our model
is the extension of the model introduced in~\cite{Pires2015} to the case of two
dimensions. We are able to deduce the continuous limit for the model, which
culminate in a diffusion equation with a singular diffusion coefficient.
We observe in the confined system the onset of two scale invariant
regimes. The first one occurs when the concentration at the
origin (the center of the confining region) reaches the critical percolation threshold. At
this point we observe scale invariance in the cluster size distribution as well
as fractality in the perimeter of the central cluster. At more confined
regimes, when the concentration reaches the critical threshold of singular diffusion, there is another
signature of scale invariance in the jump size distribution.
\begin{figure}[t]
 \center
 \includegraphics[width=\columnwidth]{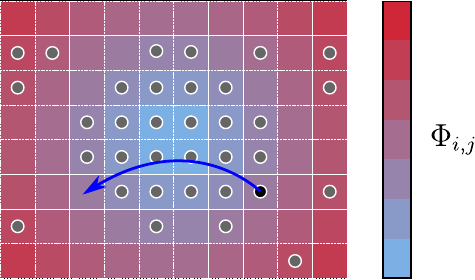}
 \caption{Schematic picture of the 2D confined sandpile model. The sites with
 discs inside are occupied, the site with a black disc inside is the source site
 of the jump, and the arrow points to the target site of the jump. The jump is
 the colored arrow. The tone gradient indicates the local
 potential $\Phi$.}\label{Fig::model}
\end{figure}

\section{Model Formulation}

\begin{figure}[t]
 \center
 \includegraphics[width=\columnwidth]{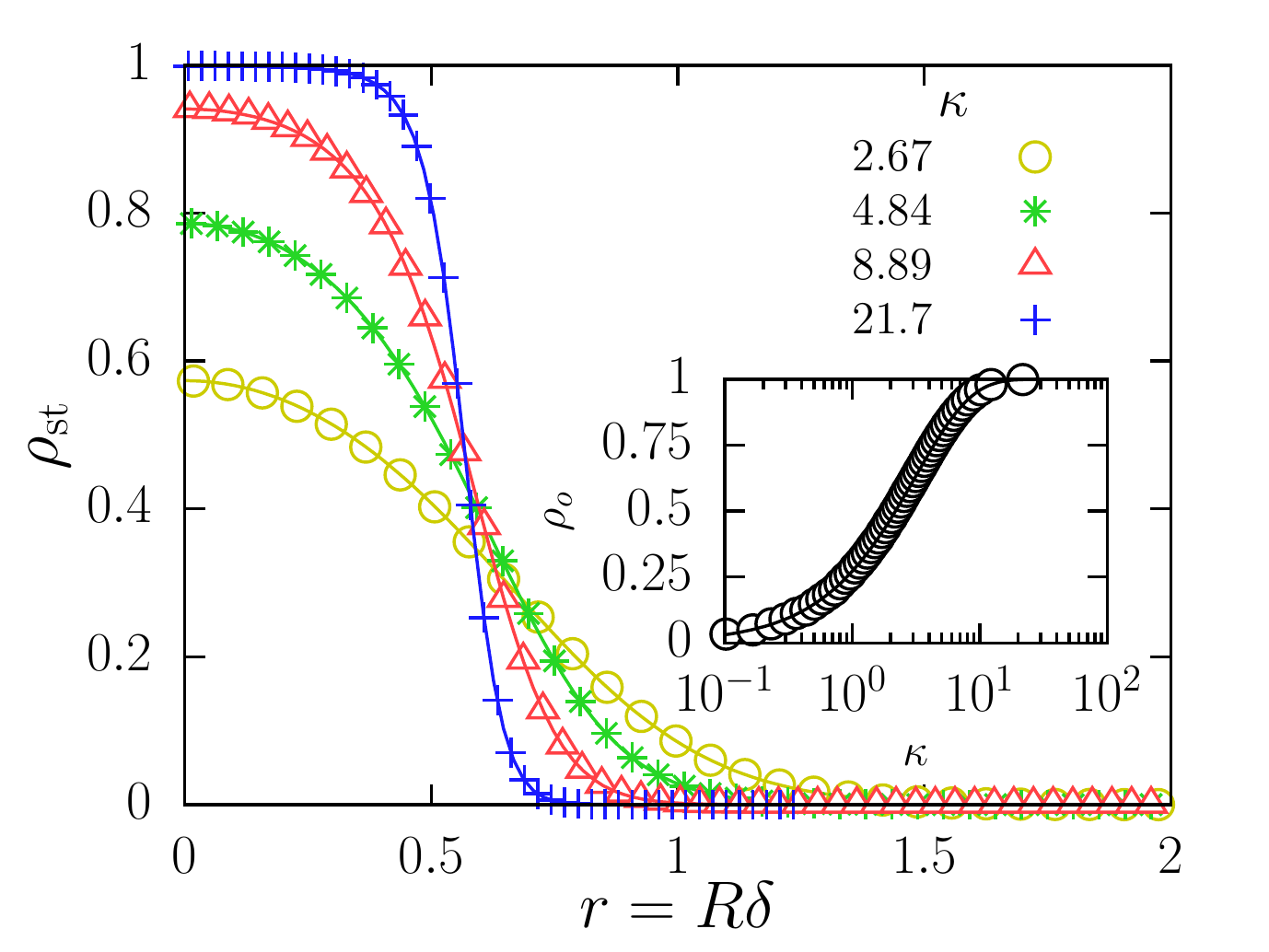}
 \caption{Occupation density as function of $r$ for different values of
 $\kappa$. The points are estimated from a radial
 histogram of the occupation averaged over time with $N_p=8000$
 particles, while the solid lines are the predictions of Eq.(\ref{Eq::rho_st2}).
 The inset shows the occupation density calculated at $r=0$. The
 solid line is the analytical prediction
 $\rho_o=1-\exp(-\beta\kappa{N}_p\delta^2/\pi)$.}\label{Fig::rho_r_po_k}
\end{figure}

\begin{figure}[b]
 \center
 \includegraphics[width=\columnwidth]{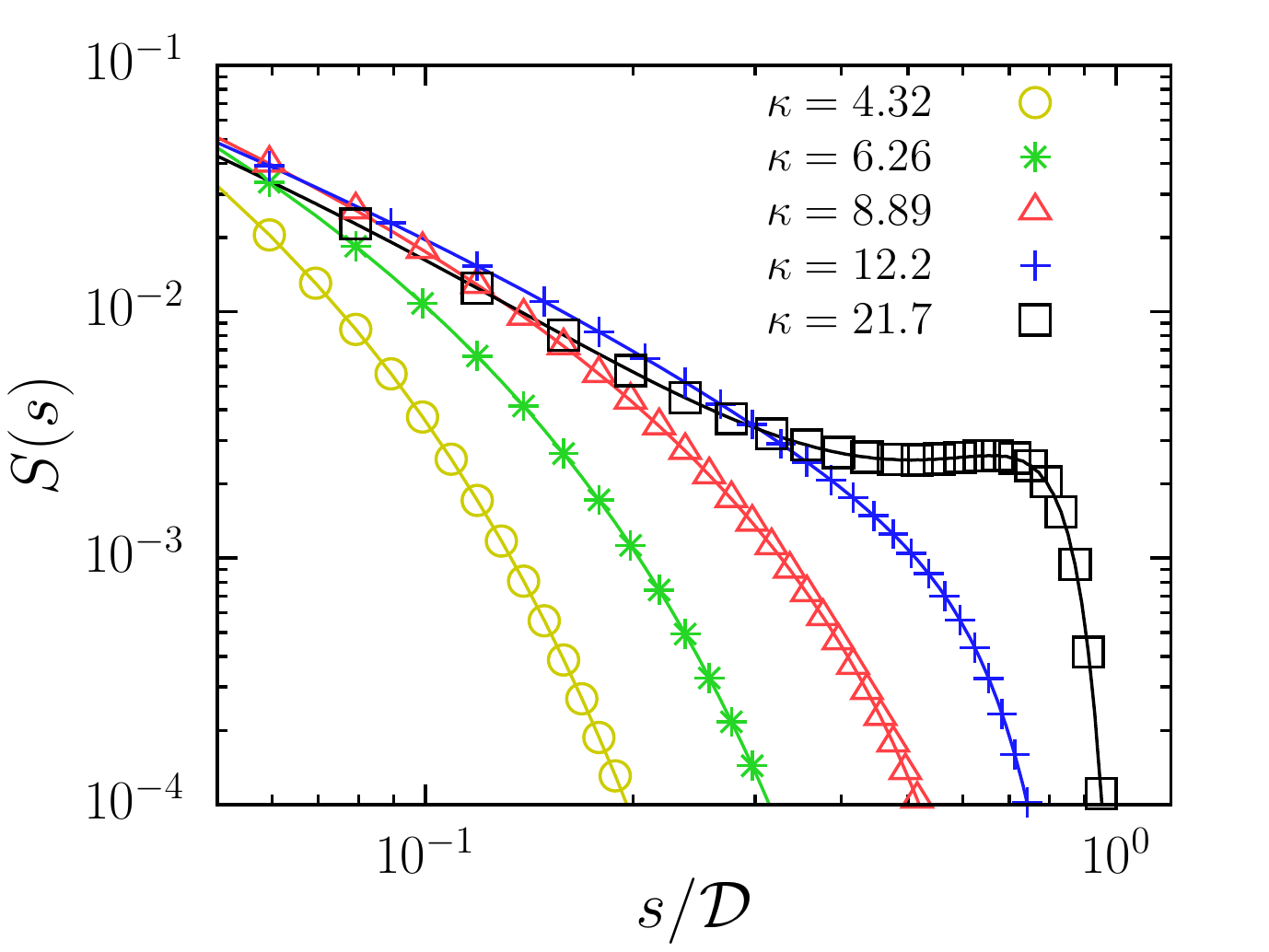}
 \caption{Jump size distribution for different values of $\kappa$ with
 $N_p=8000$. The points are numerical simulations and the solid
 lines are the predictions of Eq.~(\ref{Eq::S_s}). Here we define
 $\mathcal{D}=2\sqrt{N_p/\pi}$ as the diameter of a dense
 disc with $N_p$ particles.} \label{Fig::S_s}
\end{figure}

In the present model, $N_p$ particles are placed
on a square lattice where we define the occupation $h_{i,j}$ for each site as
$0$ (if the site is empty) or $1$ (if the site is occupied). As shown in Fig.~\ref{Fig::model}, at each iteration,
a particle can move randomly to any of four directions of the square lattice.
The particle moves from an occupied site
(the source) and get past all occupied sites on the chosen direction until it
reaches an empty site (the target) where it may stay with a given probability (see
Fig.~\ref{Fig::model}) in a movement that we label a jump. The probability that a jump is accepted is
given by the Metropolis factor $\Theta = \min(1,\exp(-\beta [\Phi_{i_t,j_t} -
 \Phi_{i_s,j_s}]))$, where we define $\Phi_{i,j}$ as the external potential
energy of a site. This probability introduces the effect of a confining potential
on the particles. Here we present result using only four directions, since, for this case, a
continuous limit of the model can be found analytically. However we also
performed simulations with a model where particles can move in any arbitrary
direction, leading to entirely similar results.

\section{Continuous limit of the model}

\begin{figure}[t]
 \center
 \includegraphics[width=\columnwidth]{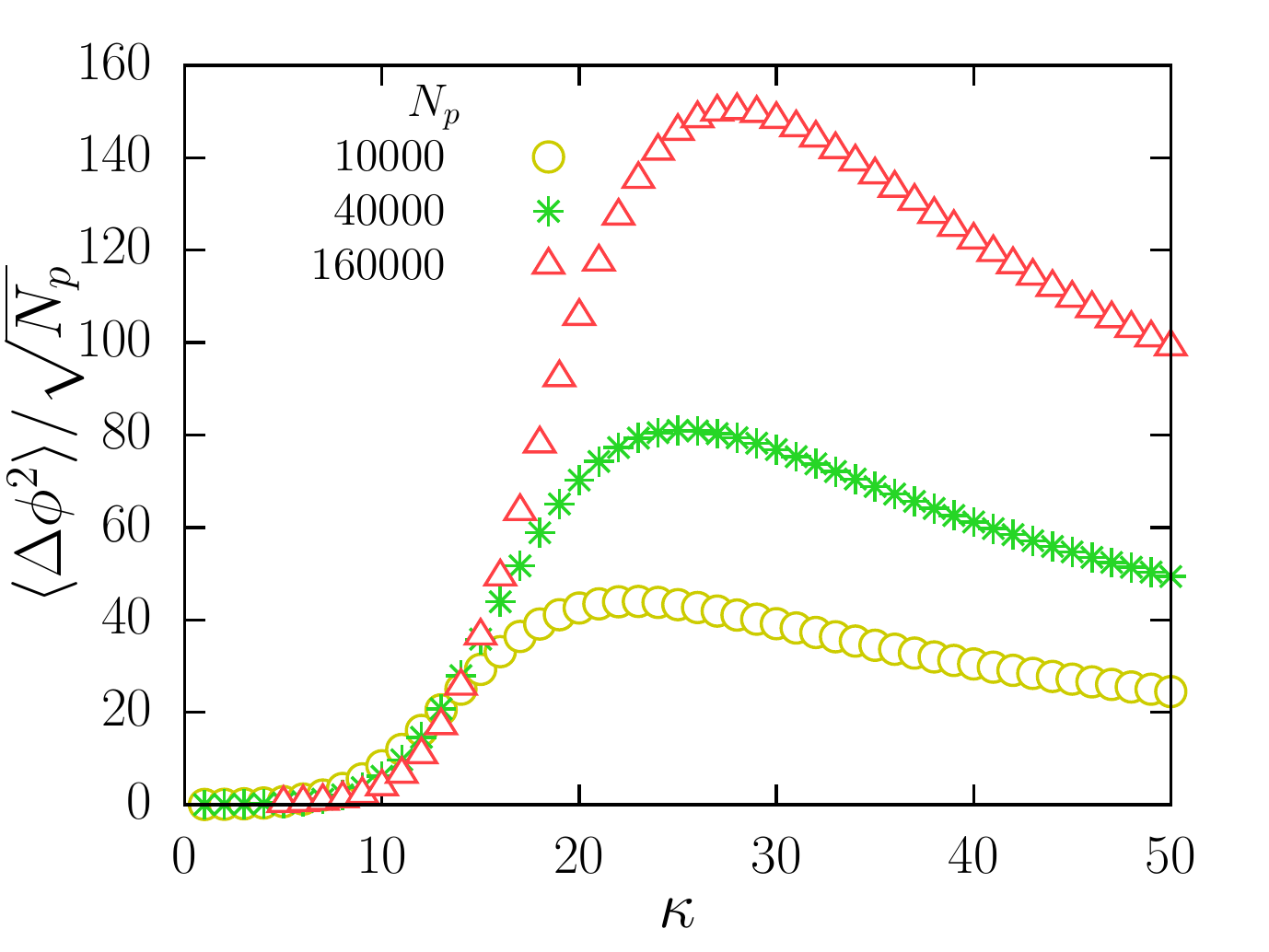}
 \caption{Mean jump energy fluctuation as function of $\kappa$ for
 different number of particles $N_p$. The points are the results
 from Eq.~\ref{Eq::msjef}. For each value of $N_p$,
 the energy fluctuation exhibits a maximum at a different value
 $\kappa^{*}(N_p)$.}\label{Fig::E2_kappa}
\end{figure}

It is possible to verify that the model we described is a Markov process,
that is, the occupation probability of a site $(i,j)$ at a given step $n+1$, $\rho_{i,j|n+1}$,
can be obtained from the occupation probability of all sites on the previous
step,

\begin{eqnarray}
	 \rho_{i,j|n+1}&=&\rho_{i,j|n} \nonumber \\
	 &+&\sum_{i^\prime\ne i}\frac{1}{4}P_{i,i^\prime-1|j}\rho_{i^\prime,j|n}
		 (1-\rho_{i,j|n})\Theta_{i,i^\prime|j} \nonumber \\
	 &-&\sum_{i^\prime\ne i}\frac{1}{4}P_{i,i^\prime-1|j}\rho_{i,j|n}
		 (1-\rho_{i^\prime,j|n})\Theta_{i^\prime,i|j} \nonumber \\
	 &+&\sum_{j^\prime\ne j}\frac{1}{4}P_{i|j,j^\prime-1}\rho_{i,j|n}
		 (1-\rho_{i,j|n})\Theta_{i|j,j^\prime} \nonumber \\
	 &-&\sum_{j^\prime\ne j}\frac{1}{4}P_{i|j,j^\prime-1}\rho_{i,j|n}
		 (1-\rho_{i,j^\prime|n})\Theta_{i|j^\prime,j}\label{Eq::stochastic},
\end{eqnarray}
where
$\Theta_{i|j,j^\prime}=\min(1,\exp(-\beta[\Phi_{i,j}-\Phi_{i,j^\prime}]))$
is the Metropolis factor, and $P_{i|j,j^\prime}$ is the probability of
finding all sites between $j$ and $j^\prime$ in the $i$ column
occupied. Similarly, $\Theta_{i,i^\prime|j} =
\min(1,\exp(-\beta[\Phi_{i,j} - \Phi_{i^\prime,j}]))$, where
$P_{i,i^\prime|j}$ is the probability of finding all sites between
column $i$ and $i^\prime$ on the $j$ line occupied. We define
the probabilities of finding $k$ consecutive
sites occupied in a given direction as
\begin{equation*}
 \Omega^{\pm}_{i,j|k}=\prod_{k^\prime = 1}^k \rho_{i\pm k^\prime, j} \quad \text{and} \quad
 \Psi^{\pm}_{i,j|k}=\prod_{k^\prime = 1}^k \rho_{i,j\pm k^\prime},
\end{equation*}
where $\Omega^{\pm}_{i,j|k}$ is used for the horizontal direction and
$\Psi^{\pm}_{i,j|k}$ is used for the vertical direction. We then define,
\begin{equation*}
 \Xi^{\pm}_{i,j}= \sum_{i^\prime=1}^{\infty}\Theta_{i,i\pm i^\prime|j}
 \Omega^{\pm}_{i,j|i^\prime}
\end{equation*}
as the contribution for the probability due to particles arriving from the
horizontal direction. In a similar fashion, we define
$\Gamma^{\pm}_{i,j}$ as the contribution due to particles arriving
from vertical direction, and
\begin{equation*}
 \Lambda^{\pm}_{i,j}= \sum_{i^\prime=1}^{\infty} (1-\rho_{i\pm i^\prime,j})
 \Theta_{i\pm i^\prime,i|j} \Omega^{\pm}_{i,j|i^\prime-1}
\end{equation*}
as the contribution due to particles leaving in the horizontal
direction. Finally, $\Upsilon^{\pm}_{i,j}$ is the contribution due to
particles leaving in vertical direction.

\begin{figure}[t]
 \center
 \includegraphics[width=\columnwidth]{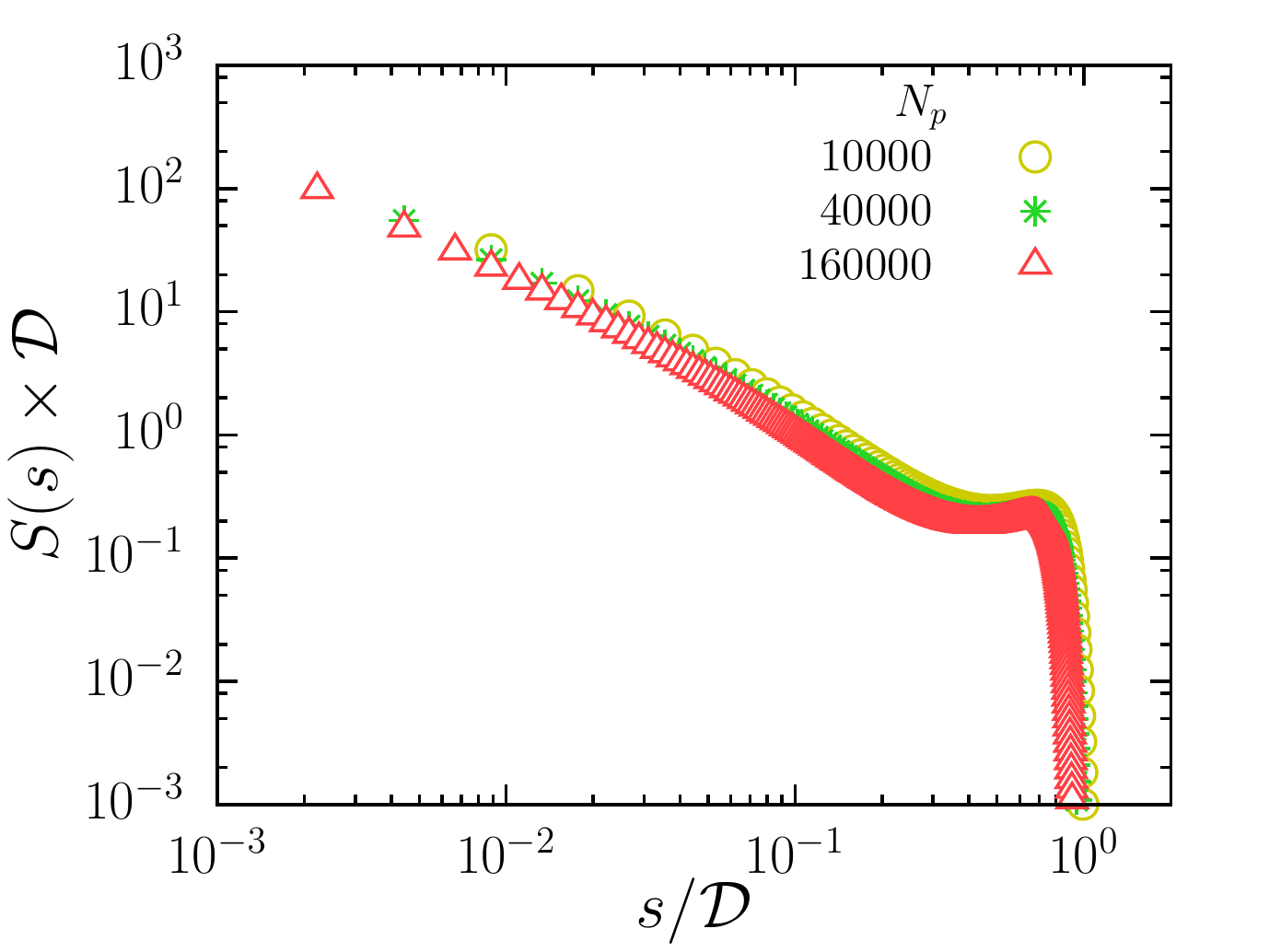}
 \caption{Jump size distribution for $\kappa=\kappa^{*}$ and
 different values of $N_p$. This result is obtained through Eq.~\ref{Eq::S_s}
 choosing $S_o$ in such a way that $S(s)$ is normalized. The factor
 $\mathcal{D}=2\sqrt{N_p/\pi}$ is the diameter of a compact cluster
 with all particles. } \label{Fig::S_s2}
\end{figure}

Using these definitions, Eq.~(\ref{Eq::stochastic}) can be written as
\begin{eqnarray}
 \frac{\rho_{i,j|n+1}-\rho_{i,j|n}}{\tau}&=&
 \frac{(1-\rho_{i,j})}{4\tau}( \Xi^{+}_{i,j} + \Xi^{-}_{i,j}
 + \Gamma^{+}_{i,j} + \Gamma^{-}_{i,j} )\nonumber \\
 &-&\frac{\rho_{i,j}}{4\tau}(
 \Lambda^{+}_{i,j} + \Lambda^{-}_{i,j} 
 + \Upsilon^{+}_{i,j} + \Upsilon^{-}_{i,j}),
 \label{Eq::master}
\end{eqnarray}
where $\tau$ is the time unit. The first factor on the
right side of Eq.~(\ref{Eq::master}) accounts for particles arriving
at a given site from each of four directions of the lattice, while the
second factor on the left side of Eq.~(\ref{Eq::master}) accounts for
particles leaving this site in each of the four directions of the
lattice.

\begin{figure}[t]
 \center
 \includegraphics[width=\columnwidth]{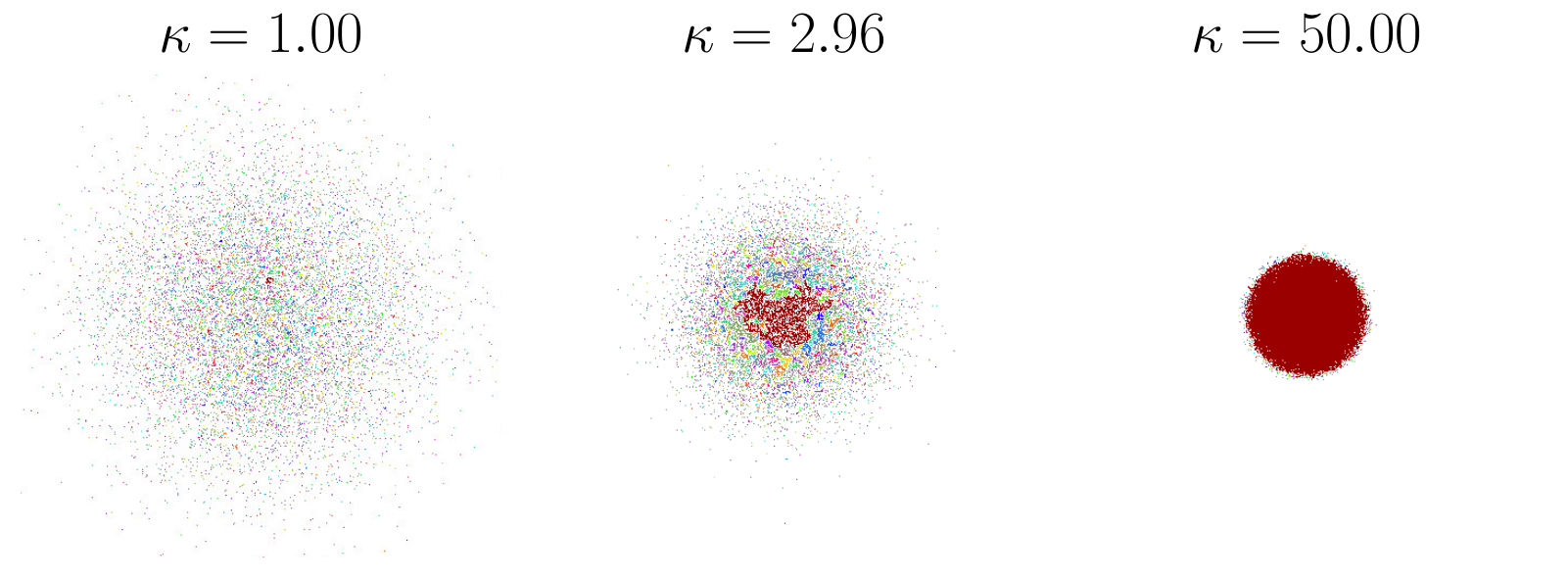}
 \caption{Snapshots of simulations of systems with 10000 particles
 for different value of $\kappa$. The colors indicate different clusters.
 Observe the relative size of the clusters and the formation of a giant cluster
 when $\kappa$ increases.
 }\label{Fig::clusters}
\end{figure}

The continuous limit of Eq.~(\ref{Eq::master}) can be obtained similarly to what
was done for 1D~\cite{Pires2015}, resulting in the following non-linear
diffusion equation
\begin{equation}
 \frac{\partial \rho}{\partial t} = D \nabla \cdot \left[
 \frac{(1+\rho)}{(1-\rho)^3}\nabla\rho +
 \beta\frac{(1+\rho)}{(1-\rho)^2}\rho\nabla\Phi \right]\label{Eq::diffusion},
\end{equation}
where we define
\begin{equation}
D=\mathop{\lim_{\mathop{\delta \rightarrow 0}}}_{\tau
\rightarrow 0}\frac{\delta^2}{4\tau},\label{Eq::D}
\end{equation}
with $\delta$ been the space unit used for the space continuous limit and
$\rho=\rho(x=i\delta,y=j\delta,t=n\tau)$. From Eq.~\ref{Eq::diffusion},
we see that our model obeys the continuity equation
$\partial \rho / \partial t = - \nabla \mathbf{J}$, where we have
\begin{equation*}
 \mathbf{J} = -D \left[
 \frac{(1+\rho)}{(1-\rho)^3}\nabla\rho +
 \beta\frac{(1+\rho)}{(1-\rho)^2}\rho\nabla\Phi \right].
\end{equation*}
In the case where the dependence of the potential is only radial,
$\Phi=\Phi(r)$, the conditions for a stationary solution are
$\partial \Phi/\partial r = 0$, $\rho_{\text{st}}(r\rightarrow\infty)\rightarrow 0$
and $\mathbf{J}=\mathbf{0}$. Thus,
\begin{equation*}
 r\left[\frac{(1+\rho)}{(1-\rho)^3}\frac{d\rho}{dr} +
 \beta\frac{(1+\rho)}{(1-\rho)^2}\rho\frac{d\Phi}{dr} \right] = 0.
\end{equation*}
which can be written as
\begin{equation}
 \frac{1}{\beta(1-\rho)\rho}\frac{d\rho_{\text{st}}}{dr}=-\frac{d\Phi}{dr}
 =\frac{d}{dr}\left[\int_{u_0}^{\rho_{\text{st}}}\frac{1}{\beta(1-u)u}du\right]\label{Eq::dphi_dr}.
\end{equation}
The Eq.~(\ref{Eq::dphi_dr}) can be easily solved and results in the stationary solution
given by
\begin{equation}
 \rho_{\text{st}}(r) = \frac{1}{1+e^{\beta[\Phi(r)-\mu]}}, \label{Eq::rho_st}
\end{equation}
where $\mu$ can be obtained from the constraint $\int \frac{\rho(r)}{\delta^2}dS
= N_p$, leading to
\begin{equation*}
 \int_0^\infty \frac{1}{1+e^{\beta[\Phi(r)-\mu]}}rdr = \frac{N_p\delta^2}{2\pi}.
\end{equation*}

In the particular case of a parabolic confining potential,
$\Phi(r=R\delta)=\kappa r^2=\kappa (i^2+j^2)\delta^2$, this integral can be
solved, and it is possible to show that $\mu=(1/\beta)\ln(\exp(\beta\kappa
N_p\delta^2/\pi) - 1)$, so that Eq.~(\ref{Eq::rho_st}) is now independent of
$\mu$
\begin{equation}
 \rho_{\text{st}}(r) = \frac{(e^{\beta\frac{\kappa N_p\delta^2}{\pi}} - 1)}
 {e^{\beta\frac{\kappa N_p\delta^2}{\pi}} + (e^{\beta \kappa r^2} - 1)}.
 \label{Eq::rho_st2}
\end{equation}

\begin{figure}[t]
 \center
 \includegraphics[width=\columnwidth]{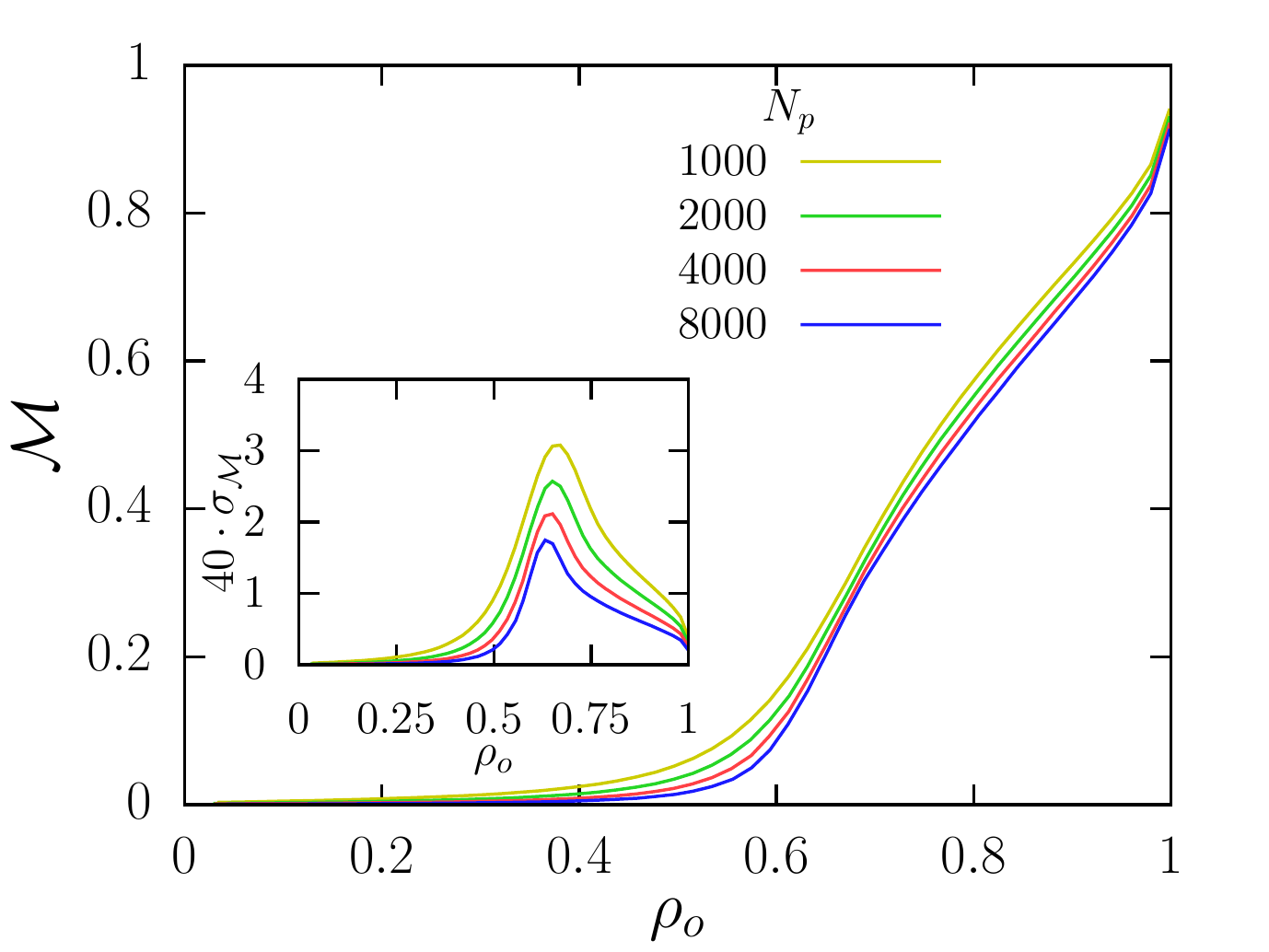}
 \caption{
 Fraction of the largest cluster $\mathcal{M}$ computed from numerical simulations as
 function of occupation density at the origin $\rho_o$ for different
 values of number of particles $N_p$. The inset shows the standard deviation
 of $\mathcal{M}$ as function of $\rho_o$.} \label{Fig::M_rho}
\end{figure}

In Fig.~\ref{Fig::rho_r_po_k}, we can see the agreement between
predictions from Eq.~(\ref{Eq::rho_st2}) and results from numerical
simulations. Due to the intrinsic exclusion mechanism of the
model, the stationary state given by Eq.~(\ref{Eq::rho_st}) is a
Fermi-Dirac distribution~\cite{Fernando2010}, and as $\kappa$ increases, the occupation
tends to saturate at $\rho_{\text{st}}=1$ near $r=0$. This behavior
leads to the formation of a giant cluster of particles near the origin
 as $\kappa$ increases. The occupation at $r=0$
follows $\rho_o=1-\exp(-\beta\kappa N_p\delta^2/\pi)$, and, as the
inset in Fig.~\ref{Fig::rho_r_po_k} shows, this prediction follows
closely the results from numerical simulations.

\section{Jump Size Distribution and Mean Square Energy Fluctuation}

At this point, it is important to define a quantity that can give
information about the dynamics of the model, which is the jump size
distribution

\begin{equation*}
 S(s)\approx \lim_{t\rightarrow \infty}\left[ \frac{\mathcal{N}_{[t]}(s)}{\mathcal{N}_{[t]T}} \right ],
\end{equation*}
where $s$ is the size of a jump, $S(s)$ is the probability of a jump
of size $s$ to occur, $\mathcal{N}_{[t]}(s)$ is the number of jumps
of size $s$ that appeared in a time interval $t$, and
$\mathcal{N}_{[t]T}$ is the total number of jumps in a time
interval. Assuming that it is possible to obtain the dynamics of the model from the
occupation probability, we can use the following approximation:

\small
\begin{equation}
 S(s)\equiv S_o\sum_{i}\sum_{j} \frac{{\rho}_{i,j}}{4}
 \left[\lambda^{+}_{i,j|s}+\lambda^{-}_{i,j|s}
 +\upsilon^{+}_{i,j|s}+\upsilon^{-}_{i,j|s}\right],\label{Eq::S_s}
\end{equation}
\normalsize
where $S_o$ is a normalization constant, and we obtain the
occupation probability from the stationary condition,
$\rho_{i,j}=\rho_{st}(x=i\delta,y=j\delta)$. The factors
$\lambda^{\pm}_{i,j|s}$ account for the chance of a jump to be accepted in
the horizontal direction and is defined as
\begin{equation*}
 \lambda^{\pm}_{i,j|s}= (1-\rho_{i\pm s,j}) \Theta_{i\pm s,i|j}
 {\Omega}^{\pm}_{i,j|s-1}.
\end{equation*}
Similarly, $\upsilon^{\pm}_{i,j|s}$ accounts for the chance of a jump being accepted
in the vertical direction and is defined in a similar fashion.
We performed the summations of Eq.(\ref{Eq::S_s}) numerically and compared
it with the results from numerical simulations. The results are displayed
in Fig.~\ref{Fig::S_s}, showing very good agreement.

\begin{figure}[t]
 \center
 \includegraphics[width=\columnwidth]{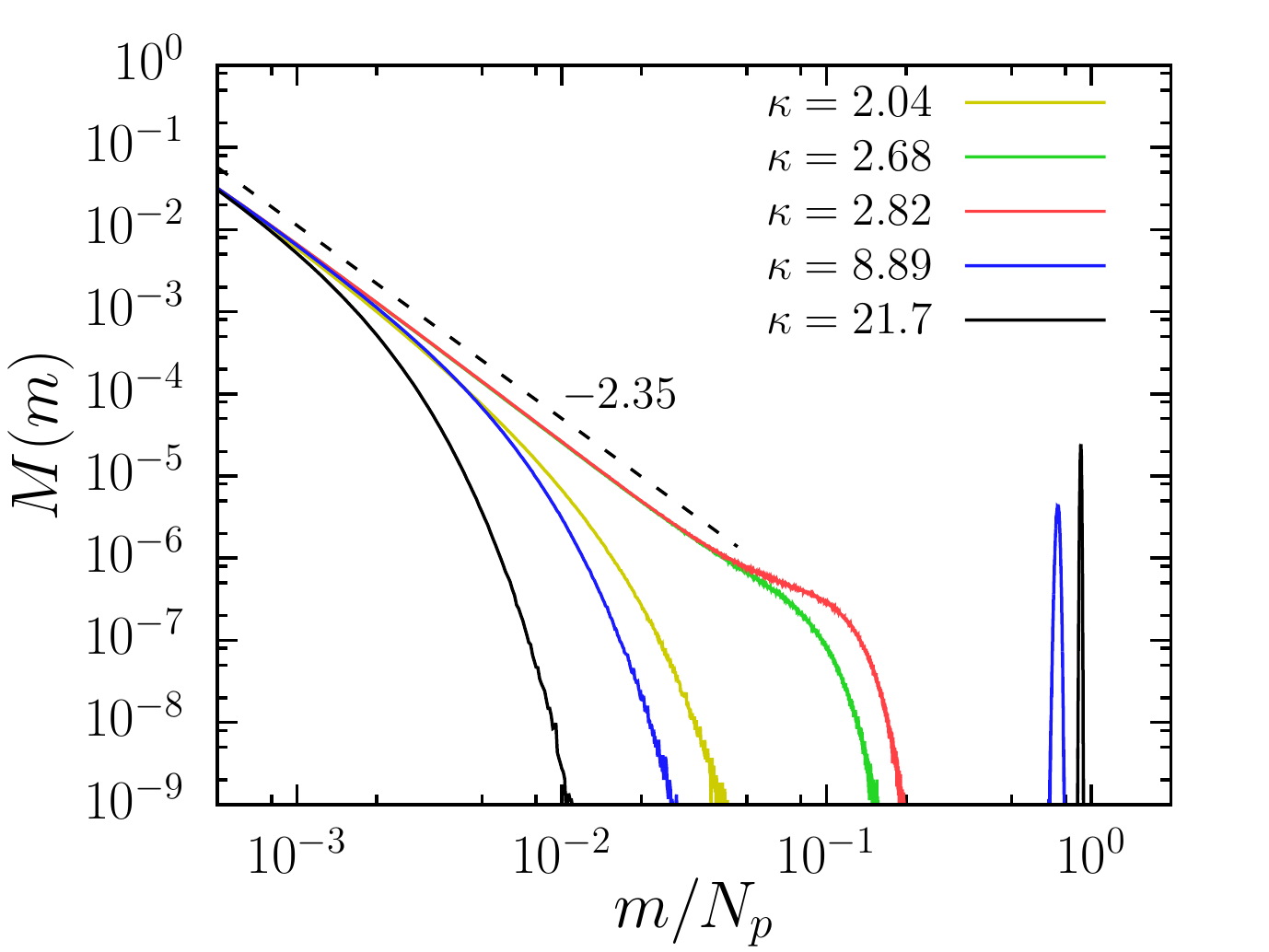}
 \caption{
 Cluster size distribution obtained through numerical simulations for $N_p=8000$
 and different values of $\kappa$. The dashed line is a power law with exponent
 equal to $-2.35$.} \label{Fig::M_m}
\end{figure}

\begin{figure}[b]
 \center
 \includegraphics[width=\columnwidth]{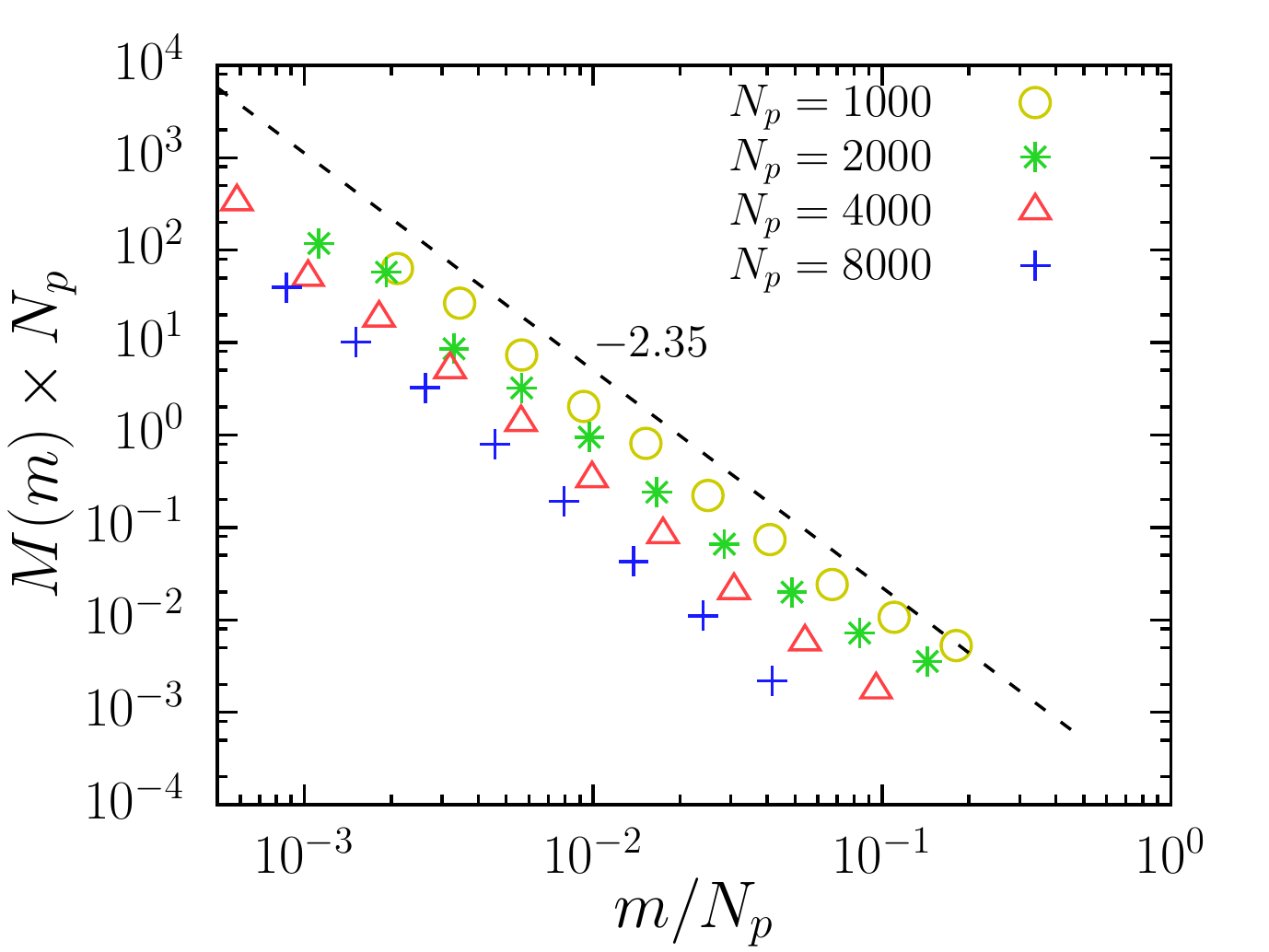}
 \caption{
 Log-log plot of the cluster size distribution obtained through numerical simulations for different
 values of $N_p$ and $\kappa=\kappa^\prime$. The dashed line is a power law with
 exponent equal to $-2.35$.
 } \label{Fig::M_m2}
\end{figure}

\begin{figure}[t]
 \center
 \includegraphics[width=\columnwidth]{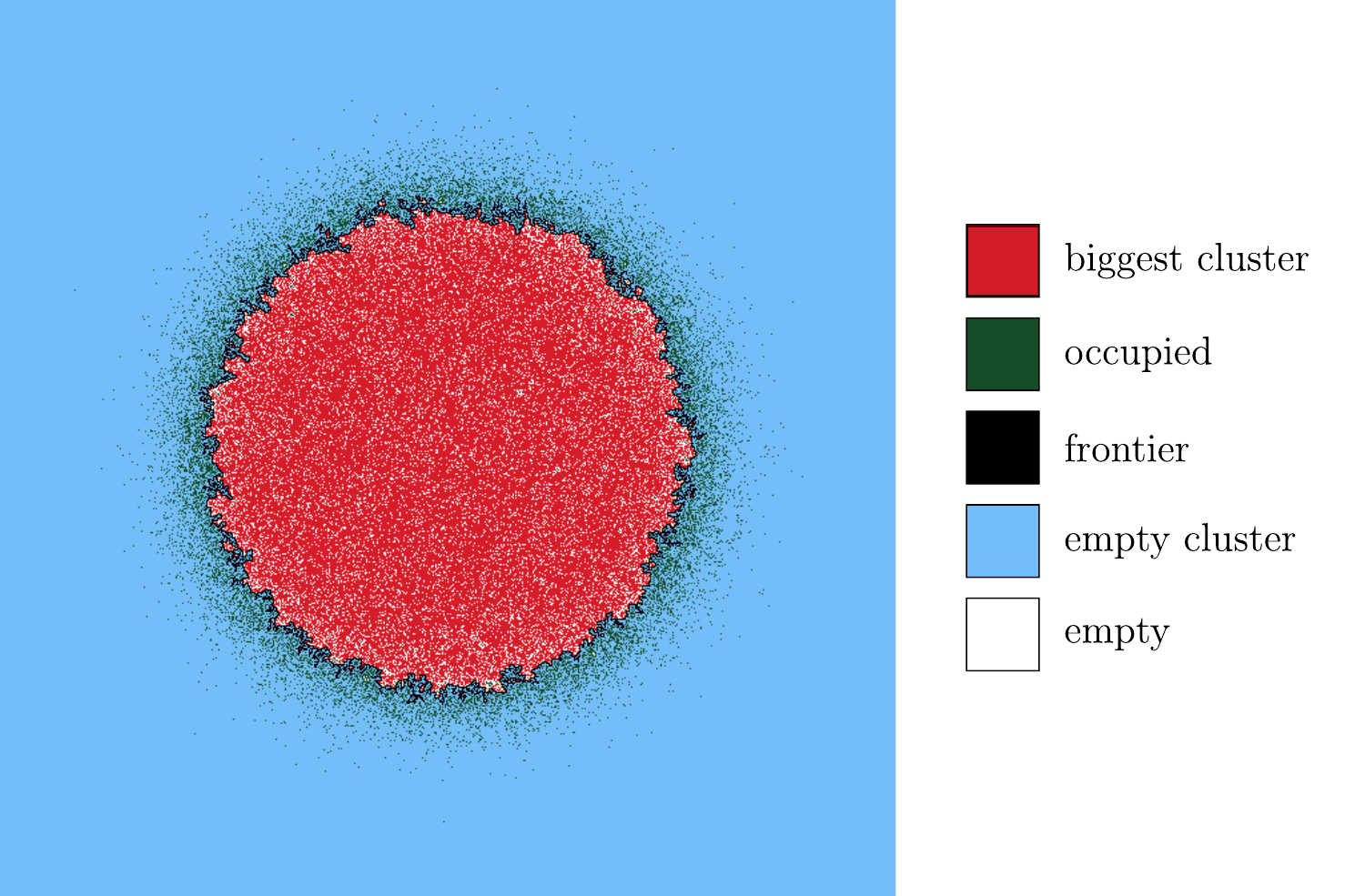}
 \caption{Diffusion frontier with $N_p=100000$ and $\kappa=20.0$. This
 picture shows the results of a simulation on a 2D square lattice were we put $100000$ particles following the
 probability given by Eq.~(\ref{Eq::rho_st2}). The red sites are the occupied
 sites that belong to the largest cluster and the black are the largest cluster
 sites that belong to the diffusion frontier. The gray sites are the
 occupied sites that do not belong to the largest cluster. The white sites are
 empty, and the blue ones correspond to the empty sites that belong to the empty cluster
 (connected through first or second neighbors to the more external empty sites).
 In this picture, there are 5960 sites at the diffusion frontier.
 }\label{Fig::hull}
\end{figure}

As $\kappa$ increases, the occupation at the center of the system saturates to
one (see Fig.~\ref{Fig::rho_r_po_k}). Consequently, in very confined systems,
jumps that pass through the whole system have larger probability, as indicated
by the peak at a large value of $s$ in Fig.~\ref{Fig::S_s} when $\kappa=21.7$.
The agreement between numerical simulations and the predictions from
Eq.~(\ref{Eq::S_s}) shown in Fig.~\ref{Fig::S_s} allow us to
confidently study larger systems and extract useful information
without the need to run time expensive simulations.

Another useful information about the dynamics of the model is the mean
square energy fluctuation of the system, which can be determined from the probabilities 
$\lambda^{+}_{i,j|s}$ and $\upsilon^{+}_{i,j|s}$ as,
\begin{eqnarray}
 \langle\Delta\phi^2\rangle&\equiv& \sum_{s}\sum_{i}\sum_{j}
 \frac{\rho_{i,j}}{4} \left[\lambda^{+}_{i,j|s}
 (\Phi_{i,j+s}-\Phi_{i,j})^2 \right. \nonumber\\
 &+&\lambda^{-}_{i,j|s} (\Phi_{i,j-s}-\Phi_{i,j})^2 +\upsilon^{+}_{i,j|s}
 (\Phi_{i+s,j}-\Phi_{i,j})^2 \nonumber\\ &+&\left.\upsilon^{-}_{i,j|s}
 (\Phi_{i-s,j}-\Phi_{i,j})^2 
 \right].\label{Eq::msjef}
\end{eqnarray}
where the sum goes over all jump sizes $s$ and all sites $(i,j)$.
Figure~\ref{Fig::E2_kappa} shows the $\langle\Delta\phi^2\rangle$ as
function of $\kappa$ for different number of particles. As can be seen,
this function have a maximum in a specific value of $\kappa^{*}$ for
each value of $N_p$. The shape of the jump size distribution at the
condition of maximum energy fluctuation appears to be size invariant,
as shown in Fig.~\ref{Fig::S_s2}. The distributions in
Fig.~\ref{Fig::S_s2} collapse when the jump size is scaled by the diameter
$\mathcal{D}=2\sqrt{N_p/\pi}$ of a dense cluster with all
particles. This result is suggestive of a scale invariant regime
observed at $\kappa=\kappa^{*}$.

\begin{figure}[t]
 \center
 \includegraphics[width=\columnwidth]{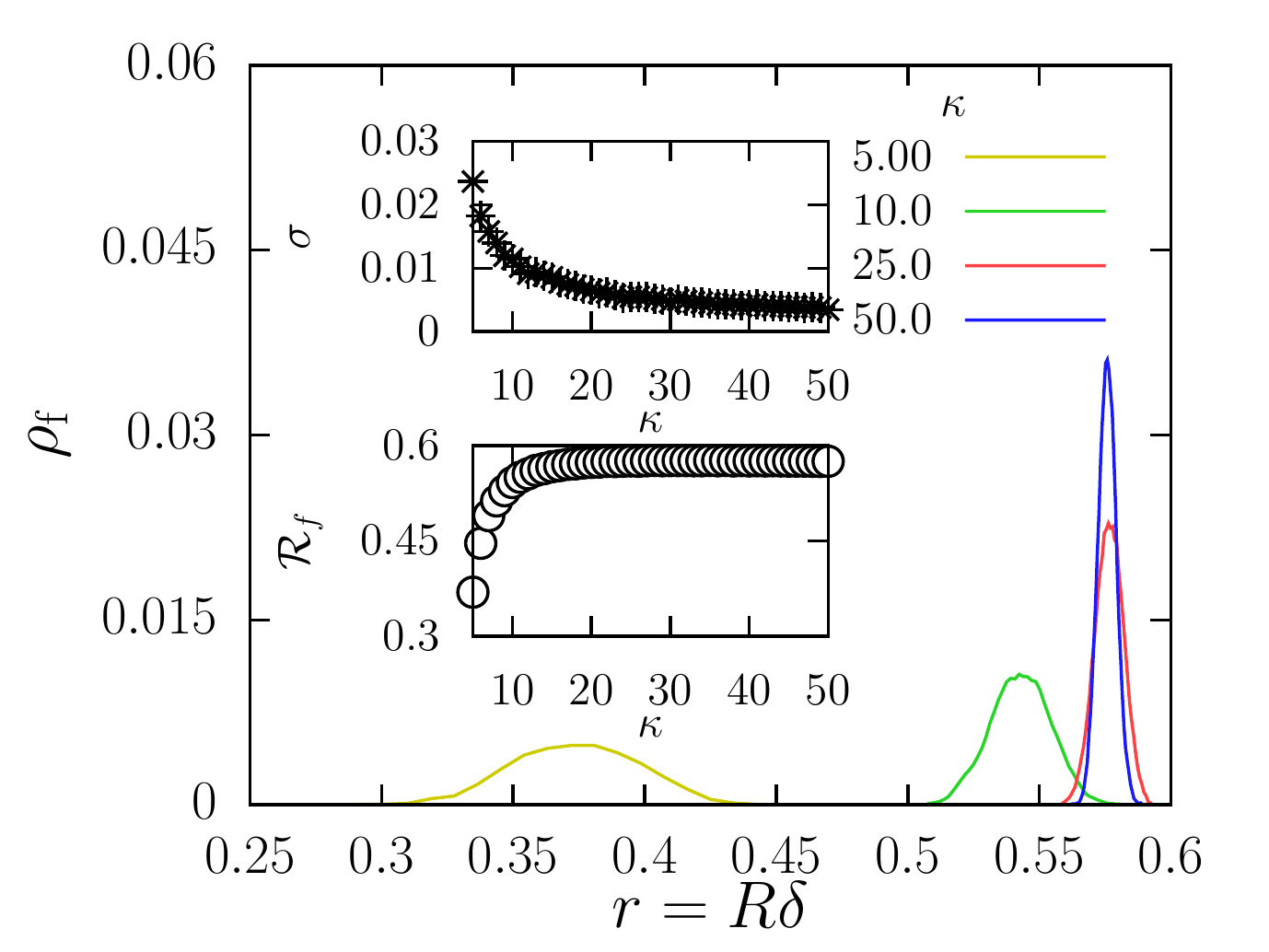}
 \caption{Probability of finding a diffusion frontier site at a
 distance $r$. In the inset, we show $\sigma$, the standard deviation of the
 distance of the diffusion frontier sites to origin, and
 $\mathcal{R}_f$, the diffusion frontier radius as function of the
 strength $\kappa$ of the confining potential. Here we used $N_p=10000000$.
 }\label{Fig::Rf}
\end{figure}

\section{Cluster Size Distribution and Diffusion Frontier}

In this section we investigate some of the geometric aspects of our model.
Figure~\ref{Fig::clusters} shows the formation of a giant cluster centered at
the origin of the confining potential, as the strength $\kappa$ increases. As
$\kappa$ grows, there is a clear percolation-like behavior induced where the
large cluster starts to grow at the origin. This can be observed quantitatively
by investigating the mean largest cluster size $\mathcal{M}$ which can be
obtained at each time step. Figure~\ref{Fig::M_rho} shows how the mean largest
cluster size changes when $\kappa$ increases. For convenience, we plot
$\mathcal{M}$ as a function of the density at the origin $\rho_o$, which
increases monotonically with $\kappa$, (see the inset of
Fig.~\ref{Fig::rho_r_po_k}). As depicted, there is a sudden increase in the size
of the largest cluster for $\rho_o\approx p_c$, where $p_c=0.59274621(13)$ is
the percolation critical point~\cite{Newman2000}. Figure~\ref{Fig::M_m} suggests
that this increasing in the size of the largest cluster takes place due to a
percolation-like transition that changes the cluster size distribution of this
model. This conjecture may be supported by investigating the cluster size
distribution at the condition where the confining potential starts to induce the
formation of a larger cluster in the center of the system,
$\kappa^\prime=\kappa(\rho_o=p_c)\approx 2.82$. Figure~\ref{Fig::M_m2} 
shows that the cluster size distribution at this value of $\kappa$ follows a
power law.

\begin{figure}[t]
 \center
 \includegraphics[width=\columnwidth]{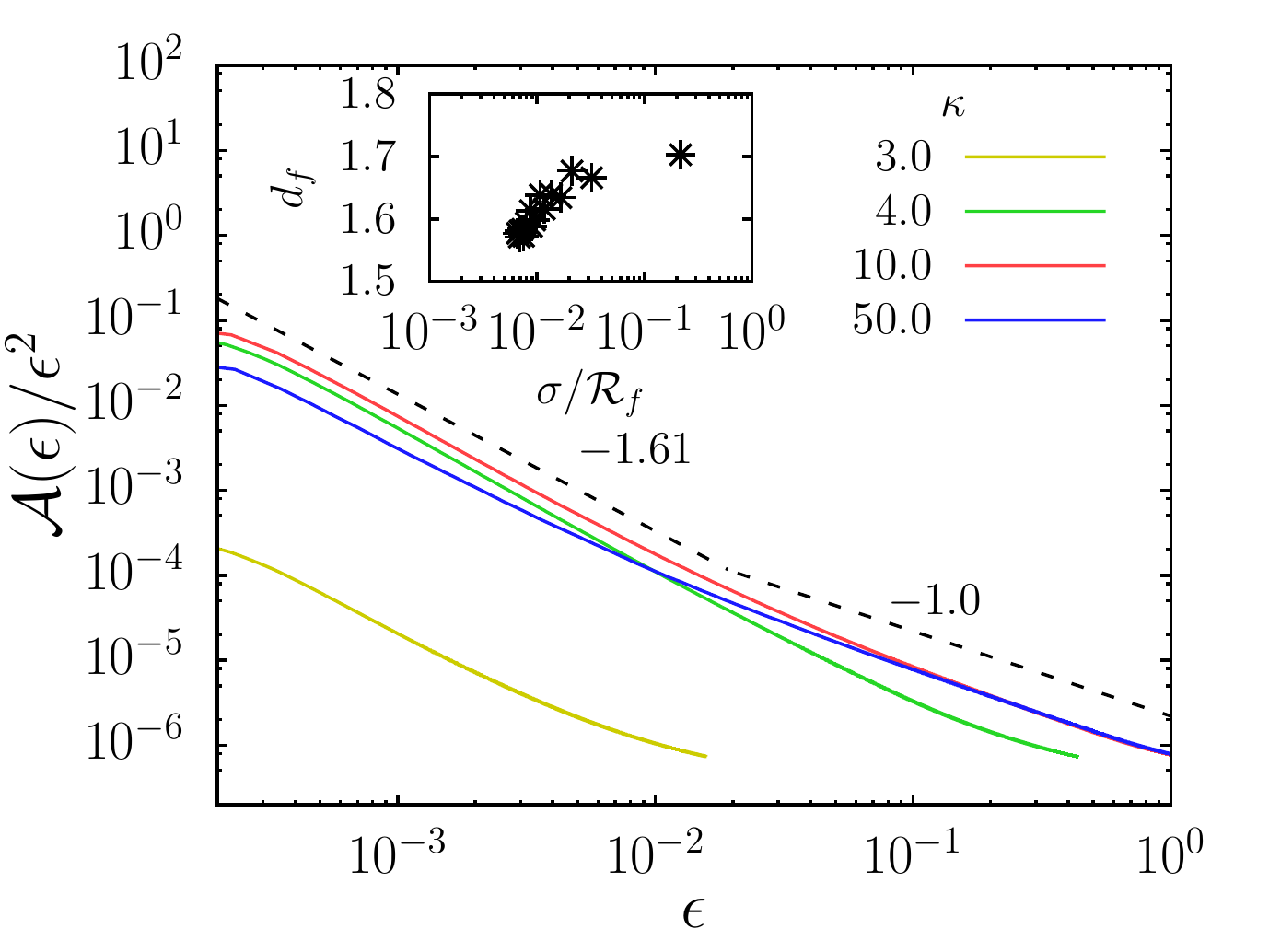}
 \caption{
 Fractal dimension determination from multiple resolution analysis. The slope of
 the dashed line is $-d_f$ according to Eq.~\ref{Eq::A_e}. The dashed line
 represents approximately the slope for most curves analyzed. The inset shows us
 how $d_f$ grows with $\sigma/\mathcal{R}_f$.
 } \label{Fig::A_e}
\end{figure}

Another relevant quantity that can be considered is the cluster external
perimeter (the so called hull) of the largest cluster. As shown in
Fig.~\ref{Fig::hull}, this structure is what we call the diffusion
frontier and its characteristics are closely related to the gradient percolation
diffusion front~\cite{Sapoval1985}. We define the frontier radius
$\mathcal{R}_f$ as the mean distance of the diffusion frontier sites to
the origin. To study systems with a lager number of particles, we used random
samples~\cite{Knuth1981} to generate a sample system from Eq.~\ref{Eq::rho_st2}.
Figure~\ref{Fig::Rf} shows the probability of finding a diffusion
frontier site at a given distance to the origin. The insets in the
Fig.~\ref{Fig::Rf} shows $\sigma$, the standard deviation of the distance of the
diffusion frontier sites to origin, and $\mathcal{R}_f$, the
diffusion frontier radius as a function of the strength $\kappa$ of the
confining potential. One important difference between our model and usual
gradient percolation models is the relation between the linear size
$\mathcal{R}_f$ and $\sigma$, since both are function of $\kappa$, in such way
that we cannot fix one and vary the other. This condition limits the values of
$\mathcal{R}_f$ and $\sigma$ that are close to the condition where finite size
effects affect the diffusion frontier. In Fig.~\ref{Fig::A_e}, we use
multiple resolution analysis~\cite{Peleg1984,Peli1990} to estimate the fractal
dimension $d_f$ of the diffusion frontier. This method consists in
considering all points at a distance less than a given $\epsilon$ from the set
for which the fractal dimension is being estimated. These points form a new set
with the area given by the Richardson law~\cite{Peli1990}
\begin{equation}
\mathcal{A}(\epsilon)\propto \epsilon^{2-d_f}.\label{Eq::A_e}
\end{equation}
One can see, at higher scales, that the frontier appears as a one dimensional line,
crossing over to a higher fractal dimension $d_f$ at smaller scales. As $\kappa$
decreases, approaching the value $\kappa^\prime$ where $\rho_0\approx p_c$, the
values of the higher dimension $d_f$ grows, but the scaling region decreases. In
the inset of Fig.~\ref{Fig::A_e}, we can see how $d_f$ changes with
$\sigma/\mathcal{R}_f$.

\section{Conclusions}

We studied a 2D two-state confined sandpile model. From a
microscopic dynamics we derived a singular diffusion equation and were able to
obtain an analytical stationary solution for the particular case of parabolic potential.
This system appears to display two scale-invariant regimes. The first is
observed when the concentration at the origin approaches
the critical value for percolation. This regime is similar to what is observed
for gradient percolation, that is, power laws in the cluster size
distribution are observed, as well as a fractal shape for the singular diffusion frontier.
The second regime is associated with more intense confinements, when the concentration in the center approaches the
maximum value, and a scale-invariant behavior is observed for the jump size
distribution. We derived an analytical expression for the jump size distribution
and find it to be in good agreement with our numerical solutions. We could,
also, find a natural way to define the onset of the scale-invariant regime as
the situation where the energy fluctuations are maximum.

\section{Acknowledgments}

We thank the Brazilian agencies CNPq, CAPES, FUNCAP, and the National Institute of
Science and Technology for Complex Systems (INCT-SC) in Brazil for financial support.


\end{document}